\documentclass[12pt]{article}
\usepackage[utf8]{inputenc}
\usepackage{fancyhdr}
\usepackage{url}
\usepackage[shortcuts]{extdash}
\usepackage{graphicx}

 {\fancyhead{}
\fancyfoot[c]{\small{\rule{17.5cm}{1pt}}}}

\begin{document}
\title{\vspace{-2.5cm}
\begin{center}
\textbf{\small{WORKSHOP REPORT}}\\\vspace{-0.5cm} \rule{17.5cm}{1pt}
\end{center}
\vspace{1cm}\textbf{Report on the 8th International Workshop on Bibliometric-enhanced Information Retrieval (BIR 2019)}}

\author{Guillaume Cabanac \\
       University of Toulouse, France\\
       \emph{guillaume.cabanac@univ-tlse3.fr}
       \and 
       Ingo Frommholz \\
       University of Bedfordshire, UK\\
       \emph{ifrommholz@acm.org} \and
      Philipp Mayr \\
       GESIS -- Leibniz Institute for the Social Sciences, Germany \\
       \emph{philipp.mayr@gesis.org} \\
       \date{}}

\maketitle \thispagestyle{fancy} 

\begin{abstract}
The Bibliometric-enhanced Information Retrieval workshop series (BIR) at ECIR tackled issues related to academic search, at the crossroads between Information Retrieval and Bibliometrics.  BIR is a hot topic investigated by both academia (e.g., ArnetMiner, CiteSeer$^\chi$, Doc\-Ear) and the industry (e.g., Google Scholar, Microsoft Academic Search, Semantic Scholar). This report presents the 8th iteration of the one-day BIR workshop held at ECIR~2019 in Cologne, Germany.

\end{abstract}

\section{Introduction}
	Searching for scientific information is a long-lived information need.  In the early 1960s, Salton was already striving to enhance information retrieval by including clues inferred from bibliographic citations \cite{Salton1963}.  The development of citation indexes pioneered by Garfield~\cite{Garfield1955} proved determinant for such a research endeavour at the crossroads between the nascent fields of Information Retrieval (IR) and Bibliometrics\footnote{Bibliometrics refers to the statistical analysis of the academic literature \cite{Pritchard1969} and plays a key role in scientometrics: the quantitative study of science and innovation~\cite{LeydesdorffAndMilojevic2015}.}.  The pioneers who established these fields in Information Science---such as Salton and Garfield---were followed by scientists who specialised in one of these \cite{WhiteAndMcCain1998}, leading to the two loosely connected fields we know of today.

	The purpose of the BIR workshop series founded in 2014 is to tighten up the link between IR and Bibliometrics.  We strive to get the `retrievalists' and `citationists' \cite{WhiteAndMcCain1998} active in both academia and the industry together, who are developing search engines and recommender systems such as ArnetMiner~\cite{TangEtAl2008}, CiteSeer$^\chi$~\cite{WilliamsEtAl2014}, DocEar~\cite{BeelEtAl2014}, Google Scholar~\cite{VanNoorden2014}, Microsoft Academic Search~\cite{SinhaEtAl2015}, and Semantic Scholar~\cite{Bohannon2016}, just to name a few.

	Bibliometric-enhanced IR systems must deal with the multifaceted nature of scientific information by searching for or recommending academic papers, patents \cite{Garfield1966}, venues (i.e., conferences or journals), authors, experts (e.g., peer reviewers), references (to be cited to support an argument), and datasets.  The underlying models harness relevance signals from keywords provided by authors, topics extracted from the full-texts, coauthorship networks, citation networks, and various classifications schemes of science.
  
	Bibliometric-enhanced IR is a hot topic whose recent developments made the news---see for instance the Initiative for Open Citations~\cite{Shotton2018} and the Google Dataset Search~\cite{Castelvecchi2018} launched on September 4, 2018, which give an impression of arising challenges subject to both communities.  We believe that BIR@ECIR is a much needed scientific event for the ‘retrievalists’ and ‘citationists’ to meet and join forces pushing the knowledge boundaries of IR applied to literature search and recommendation.

\section{Past Related Activities}\label{sec:past}
    The BIR workshop series was launched at ECIR in 2014 \cite{MayrEtAl2014} and it was held at ECIR each year since then \cite{MayrEtAl2015,MayrEtAl2016,MayrEtAl2017,MayrEtAl2018bir}.  As our workshop has been lying at the crossroads between IR and NLP, we also ran it as a joint workshop called BIRNDL (for Bibliometric-enhanced IR and NLP for Digital Libraries) at the JCDL \cite{CabanacEtAl2016} and SIGIR \cite{MayrEtAl2017birndl,MayrEtAl2018birndl} conferences. All workshops had a large number of participants, demonstrating the relevance of the workshop's topics. The BIR and BIRNDL workshop series gave the community the opportunity to discuss latest developments and shared tasks such as the CL-SciSumm~\cite{JaidkaEtAl2018}, which was introduced at the BIRNDL joint workshop.

	The authors of the most promising workshop papers were offered the opportunity to submit an extended version for a Special Issue for the \emph{Scientometrics} journal \cite{MayrAndScharnhorst2015a,CabanacEtAl2018a} and of the \emph{International Journal on Digital Libraries} \cite{MayrEtAl2018a}.

    The target audience of our workshop are researchers and practitioners, junior and senior, from Scientometrics as well as Information Retrieval. These could be IR researchers interested in potential new application areas for their work as well as researchers and practitioners working with, for instance, bibliometric data and interested in how IR methods can make use of such data.

\section{Objectives and Topics for BIR@ECIR 2019}
	We called for original research at the crossroads of IR and Bibliometrics.  Thirteen peer-reviewed papers were accepted\footnote{See workshop proceedings: \url{http://ceur-ws.org/Vol-2345/}.} \cite{CabanacEtAl2019}: 9~long papers, 3~short papers and 1~demo paper. These report on new approaches using bibliometric clues to enhance the search or recommendation of scientific information or significant improvements of existing techniques. Thorough quantitative studies of the various corpora to be indexed (papers, patents, networks or else) were also contributed. The papers are as follows:
	
\begin{itemize}	
\item Long papers:
\begin{itemize}
      \item An interactive visual tool for scientific literature search: Proposal and algorithmic specification~\cite{BascurEtAl2019}
      \item A searchable space with routes for querying scientific information~\cite{Fabre2019}
      \item Discovering seminal works with marker papers~\cite{HaunschildAndMarx2019}
      \item  How  do  computer scientists  use  Google  Scholar?:  A  survey  of  user  interest  in  elements  on  SERPs and author profile pages~\cite{KimEtAl2019}
     \item Feature selection and graph representation for an analysis of science fields evolution: An application to the digital library ISTEX~\cite{LamirelAndCuxac2019}
     \item  Optimal citation con-text  window  sizes  for  biomedical  retrieval~\cite{LykkeNielseEtAl2019}
     \item  Bibliometric-enhanced  arXiv:  A  data  set  for  paper-based and citation-based tasks~\cite{SaierAndFarber2019}
     \item Mining intellectual influence associations~\cite{ShahAndPudi2019}
     \item Citation metrics for legal information retrieval systems~\cite{WiggersAndVerberne2019}
  \end{itemize}
  \item Short papers:
  \begin{itemize}
    \item Finding temporal trends of scientific concepts~\cite{FarberAndJatowt2019}
    \item  A preliminary study to compare  deep  learning  with  rule-based  approaches  for  citation  classification~\cite{PerierCambyEtAl2019}
    \item Improving scientific article visibility by neural title simplification~\cite{Shvets2019}
  \end{itemize}
 \item Demo:
 \begin{itemize}
 \item Recommending  multimedia  educational  resources  on  the MOVING platform~\cite{VaglianoAndNazir2019}.
   \end{itemize}
 \end{itemize}
 
\begin{figure}\centering
	\includegraphics[width=\linewidth]{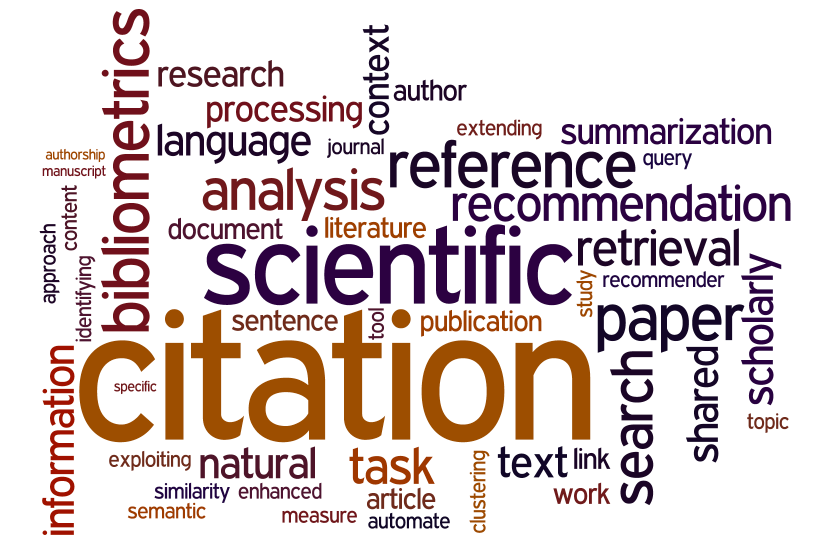}
	\caption{Main topics of the BIR and BIRNDL workshop series (2014--2018) as extracted from the titles of the papers published in the proceedings, see \protect\url{https://dblp.org/search?q=BIR.ECIR} and \protect\url{https://dblp.org/search?q=BIRNDL}.}\label{fig:wordle}
\end{figure}

    The topics of the workshop are in line with those of the past BIR and BIRNDL workshops (Fig.~\ref{fig:wordle}): a mixture of IR and Bibliometric concepts and techniques.  More specifically, the call for papers featured current research issues regarding three aspects of the search/recommendation process:
    \begin{enumerate}
        \item User needs and behaviour regarding scientific information, such as:
            \begin{itemize}
                \item Finding relevant papers/authors for a literature review;
                \item Measuring the degree of plagiarism in a paper;
                \item Identifying expert reviewers for a given submission;
                \item Flagging predatory conferences and journals.
            \end{itemize}        
        \item The characteristics of scientific information:
            \begin{itemize}
                \item Measuring the reliability of bibliographic libraries;
                \item Spotting research trends and research fronts.
            \end{itemize}
        \item Academic search/recommendation systems:
            \begin{itemize}
                \item Modelling the multifaceted nature of scientific information;
                \item Building test collections for reproducible BIR.
            \end{itemize}        
    \end{enumerate}

\section{Peer Review Process and Organization}
	The 8th BIR edition ran as a one-day workshop, as it was the case for the previous editions.  Dr.~Iana Atanassova delivered a keynote entitled ``Beyond Metadata: the New Challenges in Mining Scientific Papers'' \cite{Atanassova2019} to kick off the day.

	Two types of papers were presented: long papers (15-minute talks) and short papers (5-minute talks).  As the interactive session introduced last year was generally acclaimed, we decided to organize a interactive session to close the workshop.  Two weeks earlier, we invited all registered attendees to demonstrate their prototypes or pitch a poster during flash presentations (5 minutes).  This was an opportunity for our speakers to further discuss their work and for the public to showcase their work too.

	We ran the workshop with peer review supported by EasyChair\footnote{\url{https://easychair.org}}.  Each submission was assigned to 2 to 3 reviewers, preferably at least one expert in IR and one expert in Bibliometrics.  The stronger submissions were accepted as long papers while weaker ones were accepted as short papers, and demo.  All authors were instructed to revise their submission according to the reviewers' reports.  All accepted papers were included in the workshop proceedings \cite{CabanacEtAl2019} hosted at \url{ceur-ws.org}, an established open access repository with no author-processing charges.  

	As a follow-up of the workshop, all authors are encouraged to submit an extended version of their papers to the Special Issue of the \emph{Scientometrics} journal launched in Spring 2019.
\bibliographystyle{splncs}
\bibliography{bibdb}

\begin{thebibliography}{10}

\bibitem{Salton1963}
Salton, G.:
\newblock Associative document retrieval techniques using bibliographic
  information.
\newblock Journal of the ACM \textbf{10}(4) (1963)  440–457

\bibitem{Garfield1955}
Garfield, E.:
\newblock Citation indexes for science: {A} new dimension in documentation
  through association of ideas.
\newblock Science \textbf{122}(3159) (1955)  108--111

\bibitem{Pritchard1969}
Pritchard, A.:
\newblock Statistical bibliography or bibliometrics? [{D}ocumentation notes].
\newblock Journal of Documentation \textbf{25}(4) (1969)  348--349

\bibitem{LeydesdorffAndMilojevic2015}
Leydesdorff, L., Milojević, S.:
\newblock Scientometrics.
\newblock In Wright, J.D., ed.: {International Encyclopedia of the Social \&
  Behavioral Sciences}. Volume~21.
\newblock 2nd edn. Elsevier (2015)  322--327

\bibitem{WhiteAndMcCain1998}
White, H.D., McCain, K.W.:
\newblock Visualizing a discipline: {A}n author co-citation analysis of
  {Information Science}, 1972--1995.
\newblock Journal of the American Society for Information Science
  \textbf{49}(4) (1998)  327--355

\bibitem{TangEtAl2008}
Tang, J., Zhang, J., Yao, L., Li, J., Zhang, L., Su, Z.:
\newblock {ArnetMiner}: {E}xtraction and mining of academic social networks.
\newblock In: {KDD'08: Proceeding of the 14th ACM SIGKDD international
  conference on Knowledge discovery and data mining}, New York, NY, USA, ACM
  (2008)  990--998

\bibitem{WilliamsEtAl2014}
Williams, K., Wu, J., Choudhury, S.R., Khabsa, M., Giles, C.L.:
\newblock Scholarly big data information extraction and integration in the
  {CiteSeer}{$^\chi$} digital library.
\newblock In: {ICDE'14: Proceedings of the 30th IEEE International Conference
  on Data Engineering Workshops}, IEEE (2014)  68--73

\bibitem{BeelEtAl2014}
Beel, J., Langer, S., Gipp, B., N\"urnberger, A.:
\newblock The architecture and datasets of docear’s research paper
  recommender system.
\newblock D-Lib Magazine \textbf{20}(11/12) (2014)

\bibitem{VanNoorden2014}
Van~Noorden, R.:
\newblock {G}oogle {S}cholar pioneer on search engine’s future.
\newblock Nature (2014)

\bibitem{SinhaEtAl2015}
Sinha, A., Shen, Z., Song, Y., Ma, H., Eide, D., Hsu, B.J.P., Wang, K.:
\newblock An overview of {Microsoft Academic Service (MAS)} and applications.
\newblock In Gangemi, A., Leonardi, S., Panconesi, A., eds.: {WWW'15:
  Proceedings of the 24th International Conference on World Wide Web}, New
  York, NY, USA, ACM (2015)  243--246

\bibitem{Bohannon2016}
Bohannon, J.:
\newblock A computer program just ranked the most influential brain scientists
  of the modern era.
\newblock Science (2016)

\bibitem{Garfield1966}
Garfield, E.:
\newblock Patent citation indexing and the notions of novelty, similarity, and
  relevance.
\newblock Journal of Chemical Documentation \textbf{6}(2) (1966)  63--65

\bibitem{Shotton2018}
Shotton, D.:
\newblock Funders should mandate open citations.
\newblock Nature \textbf{553}(7687) (2018)  129

\bibitem{Castelvecchi2018}
Castelvecchi, D.:
\newblock Google unveils search engine for open data [{News \& Comment}].
\newblock Nature (2018)

\bibitem{MayrEtAl2014}
Mayr, P., Schaer, P., Scharnhorst, A., Larsen, B., Mutschke, P., eds.:
\newblock {BIR'16 Proceedings of the 1st Workshop on Bibliometric-enhanced
  Information Retrieval co-located with the 36th European Conference on
  Information Retrieval}.
\newblock Volume 1143., Aachen, CEUR-WS (2014)

\bibitem{MayrEtAl2015}
Mayr, P., Frommholz, I., Mutschke, P., eds.:
\newblock {BIR'15 Proceedings of the 2nd Workshop on Bibliometric-enhanced
  Information Retrieval co-located with the 37th European Conference on
  Information Retrieval}.
\newblock Volume 1344., Aachen, CEUR-WS (2015)

\bibitem{MayrEtAl2016}
Mayr, P., Frommholz, I., Cabanac, G., eds.:
\newblock {BIR'16 Proceedings of the 3rd Workshop on Bibliometric-enhanced
  Information Retrieval co-located with the 38th European Conference on
  Information Retrieval}.
\newblock Volume 1567., Aachen, CEUR-WS (2016)

\bibitem{MayrEtAl2017}
Mayr, P., Frommholz, I., Cabanac, G., eds.:
\newblock {BIR'17 Proceedings of the 5th Workshop on Bibliometric-enhanced
  Information Retrieval co-located with the 39th European Conference on
  Information Retrieval}.
\newblock Volume 1823., Aachen, CEUR-WS (2017)

\bibitem{MayrEtAl2018bir}
Mayr, P., Frommholz, I., Cabanac, G., eds.:
\newblock {BIR'18 Proceedings of the 7th Workshop on Bibliometric-enhanced
  Information Retrieval co-located with the 40th European Conference on
  Information Retrieval}.
\newblock Volume 2080., CEUR-WS (2018)

\bibitem{CabanacEtAl2016}
Cabanac, G., Chandrasekaran, M.K., Frommholz, I., Jaidka, K., Kan, M.Y., Mayr,
  P., Wolfram, D., eds.:
\newblock {BIRNDL'16: Proceedings of the Joint Workshop on
  Bibliometric-enhanced Information Retrieval and Natural Language Processing
  for Digital Libraries co-located with the Joint Conference on Digital
  Libraries}.
\newblock Volume 1610., Aachen, CEUR-WS (2016)

\bibitem{MayrEtAl2017birndl}
Mayr, P., Chandrasekaran, M.K., Jaidka, K., eds.:
\newblock {BIRNDL'17: Proceedings of the 2nd Joint Workshop on
  Bibliometric-enhanced Information Retrieval and Natural Language Processing
  for Digital Libraries co-located with the Joint Conference on Digital
  Libraries}.
\newblock Volume 1888., Aachen, CEUR-WS (2017)

\bibitem{MayrEtAl2018birndl}
Mayr, P., Chandrasekaran, M.K., Jaidka, K., eds.:
\newblock {BIRNDL'17: Proceedings of the 3rd Joint Workshop on
  Bibliometric-enhanced Information Retrieval and Natural Language Processing
  for Digital Libraries co-located with the Joint Conference on Digital
  Libraries}.
\newblock Volume 2132., Aachen, CEUR-WS (2018)

\bibitem{JaidkaEtAl2018}
Jaidka, K., Chandrasekaran, M.K., Rustagi, S., Kan, M.Y.:
\newblock Insights from {CL-SciSumm} 2016: {T}he faceted scientific document
  summarization shared task.
\newblock International Journal on Digital Libraries \textbf{19}(2--3) (2018)
  163--171

\bibitem{MayrAndScharnhorst2015a}
Mayr, P., Scharnhorst, A.:
\newblock Scientometrics and information retrieval: weak-links revitalized.
\newblock Scientometrics \textbf{102}(3) (2015)  2193--2199

\bibitem{CabanacEtAl2018a}
Cabanac, G., Mayr, P., Frommholz, I.:
\newblock Bibliometric-enhanced information retrieval: {P}reface.
\newblock Scientometrics \textbf{116}(2) (2018)  1225--1227

\bibitem{MayrEtAl2018a}
Mayr, P., Frommholz, I., Cabanac, G., Chandrasekaran, M.K., Jaidka, K., Kan,
  M.Y., Wolfram, D.:
\newblock Special issue on bibliometric-enhanced information retrieval and
  natural language processing for digital libraries.
\newblock International Journal on Digital Libraries \textbf{19}(2--3) (2018)
  107--111

\bibitem{CabanacEtAl2019}
Cabanac, G., Frommholz, I., Mayr, P., eds.:
\newblock {BIR'19 Proceedings of the 8th Workshop on Bibliometric-enhanced
  Information Retrieval co-located with the 41th European Conference on
  Information Retrieval}.
\newblock Volume 2345., Aachen, CEUR-WS (2019)

\bibitem{BascurEtAl2019}
Bascur, J.P., van Eck, N.J., Waltman, L.:
\newblock An interactive visual tool for scientific literature search:
  {P}roposal and algorithmic specification.
\newblock In: {Proc. of the 8th Workshop on Bibliometric-enhanced Information
  Retrieval}, CEUR-WS.org (2019)  76--87

\bibitem{Fabre2019}
Fabre, R.:
\newblock A “searchable” space with routes for querying scientific
  information.
\newblock In: {Proc. of the 8th Workshop on Bibliometric-enhanced Information
  Retrieval}, CEUR-WS.org (2019)  112--124

\bibitem{HaunschildAndMarx2019}
Haunschild, R., Marx, W.:
\newblock Discovering seminal works with marker papers.
\newblock In: {Proc. of the 8th Workshop on Bibliometric-enhanced Information
  Retrieval}, CEUR-WS.org (2019)  27--38

\bibitem{KimEtAl2019}
Kim, J., Trippas, J.R., Sanderson, M., Bao, Z., Croft, W.B.:
\newblock How do computer scientists use {G}oogle {S}cholar?: {A} survey of
  user interest in elements on {SERPs} and author profile pages.
\newblock In: {Proc. of the 8th Workshop on Bibliometric-enhanced Information
  Retrieval}, CEUR-WS.org (2019)  64--75

\bibitem{LamirelAndCuxac2019}
Lamirel, J.C., Cuxac, P.:
\newblock Feature selection and graph representation for an analysis of science
  fields evolution: {A}n application to the digital library {ISTEX}.
\newblock In: {Proc. of the 8th Workshop on Bibliometric-enhanced Information
  Retrieval}, CEUR-WS.org (2019)  88--99

\bibitem{LykkeNielseEtAl2019}
Lykke~Nielsen, B., Lavlund~Skau, S., Meier, F., Larsen, B.:
\newblock Optimal citation context window sizes for biomedical retrieval.
\newblock In: {Proc. of the 8th Workshop on Bibliometric-enhanced Information
  Retrieval}, CEUR-WS.org (2019)  51--63

\bibitem{SaierAndFarber2019}
Saier, T., F{\"a}rber, M.:
\newblock Bibliometric-enhanced {arXiv}: {A} data set for paper-based and
  citation-based tasks.
\newblock In: {Proc. of the 8th Workshop on Bibliometric-enhanced Information
  Retrieval}, CEUR-WS.org (2019)  14--26

\bibitem{ShahAndPudi2019}
Shah, T., Pudi, V.:
\newblock Mining intellectual influence associations.
\newblock In: {Proc. of the 8th Workshop on Bibliometric-enhanced Information
  Retrieval}, CEUR-WS.org (2019)  100--111

\bibitem{WiggersAndVerberne2019}
Wiggers, G., Verberne, S.:
\newblock Citation metrics for legal information retrieval systems.
\newblock In: {Proc. of the 8th Workshop on Bibliometric-enhanced Information
  Retrieval}, CEUR-WS.org (2019)  39--50

\bibitem{FarberAndJatowt2019}
F{\"a}rber, M., Jatowt, A.:
\newblock Finding temporal trends of scientific concepts.
\newblock In: {Proc. of the 8th Workshop on Bibliometric-enhanced Information
  Retrieval}, CEUR-WS.org (2019)  132--139

\bibitem{PerierCambyEtAl2019}
Perier-Camby, J., Bertin, M., Atanassova, I., Armetta, F.:
\newblock A preliminary study to compare deep learning with rule-based
  approaches for citation classification.
\newblock In: {Proc. of the 8th Workshop on Bibliometric-enhanced Information
  Retrieval}, CEUR-WS.org (2019)  125--131

\bibitem{Shvets2019}
Shvets, A.:
\newblock Improving scientific article visibility by neural title
  simplification.
\newblock In: {Proc. of the 8th Workshop on Bibliometric-enhanced Information
  Retrieval}, CEUR-WS.org (2019)  140--147

\bibitem{VaglianoAndNazir2019}
Vagliano, I., Nazir, S.:
\newblock Recommending multimedia educational resources on the {MOVING}
  platform.
\newblock In: {Proc. of the 8th Workshop on Bibliometric-enhanced Information
  Retrieval}, CEUR-WS.org (2019)  148--158

\bibitem{Atanassova2019}
Atanassova, I.:
\newblock Beyond metadata: the new challenges in mining scientific papers.
\newblock In: {Proc. of the 8th Workshop on Bibliometric-enhanced Information
  Retrieval}, CEUR-WS.org (2019)  8--13

\end{thebibliography}
\end{document}